\documentclass[twocolumn,showpacs,nature,preprintnumbers,amsmath,amssymb]{revtex4-2}
\usepackage{graphicx}
\usepackage{dcolumn}
\usepackage{bm}
\usepackage{color}
\usepackage{ulem}

\begin{document}

\preprint{Physical Review B}

\title{Isolation and manipulation of a single-donor detector in a silicon quantum dot.}

\author{$^{1,2,3}$ A.~A.~Lasek}

\author{$^1$ C.~H.~W.~Barnes}
\author{$^2$ T.~Ferrus}
\email{taf25@cam.ac.uk}
\affiliation{$^1$ Thin Film Magnetism Group, Cavendish Laboratory, J. J. Thomson Avenue, Cambridge CB3 0HE, United Kingdom}
\affiliation{$^2$ Hitachi Cambridge Laboratory, J. J. Thomson Avenue, Cambridge CB3 0HE, United Kingdom}
\affiliation{$^3$ Present address : Joint Center for Quantum Information and Computer Science, University of Maryland, United States}

\date{\today}


\begin{abstract}

We demonstrate the isolation and electrostatic control of a single phosphorus donor in a silicon quantum dot by making use of source-drain bias during cooldown and biases applied to capacitively coupled gates. Characterisation of the device at low temperatures and in magnetic fields shows single donors can be electrostatically isolated near one of the quantum dot's tunnel barriers with either  single or double occupancy. This model is well supported by capacitance-based simulations. The ability to use the $D^0$ state of such isolated donors as a charge detector is demonstrated by observing the charge stability diagram of a nearby and capacitively coupled semi-connected double quantum dot.

\end{abstract}

\maketitle

\section{Introduction}

Recent progress in lithographic techniques such as focused helium-ion-beam milling \cite{helium-ion} or extreme ultraviolet lithography \cite{EUV} as well as advances in single-ion implantation \cite{single-ion1,single-ion2} have allowed the extension of Moore's law to the atomic level. Such methods do not compromise the quality of the surfaces and interfaces, thus opening the way to new applications in high-speed computing and high-density information storage \cite{memory1a, memory1b, QIP1, QIP2}. Owing to the simplicity of their energy level structure, such atomic-scale devices are generally easier to control, easier to model and ultimately better candidates for ultralow-energy-consuming electronics compared with standard Metal-Oxide-Semiconductor (MOS) structures. Currently, such downscaling can be achieved by implanting deterministically single ions \cite{single-ion1, single-ion2} or low concentration dopants \cite{low-dopant} (non-deterministic approach) into a nanoscale architecture. However, the first requires high technical expertise and world-class equipment whereas the second is inherently random and consequently lacks scalability. Such atomic-scale devices can also be fabricated by depositing single atoms with nanometre precision using the tip of a scanning tunnelling microscope \cite{STM}. In this case, requirements for the mechanical stability of the tip imply either a long processing time or the purchase of expensive corrective and stabilisation software \cite{Zyvex1,Zyvex2}. However, the need for both a cost-effective, fast and reliable solution is still required for carrying out experimental investigations to devise atom-based applications.

Here we show that single-donor structures can be realised in doped quantum dots by taking advantage of the tunability of the electron tunnelling rates as well as the interplay, at low temperatures, between disorder conferred by randomness in dopant distribution and electron-electron interaction originating from the high doping concentration \cite{glass-behaviour1, glass-behaviour2, glass-behaviour3, glass-behaviour4}. For this purpose, phosphorous-doped silicon quantum dots offer ideal conditions for experimentation. We first demonstrate how to electrostatically isolate a single donor from the large ensemble of dopants before investigating the system characteristics and dynamics under gate voltages. We finally show this device can be used as a charge detector by tuning the detection level to the singly occupied D$^0$ state and sensing the charge occupancy of a nearby capacitively coupled double quantum dot \cite{IDQD}.

\section{Main}

\subsection{Methodology and isolation of single donors}

Most of nanoscale silicon transistors currently used in industries make use of a metal-oxide-semiconductor structure which allows controlling the number of electrons down to the single electron regime with great accuracy. Fin-field effect transistors (FinFETs) are one of many successful examples of such nano-engineering. Further downscaling to sub-10 nm structures has been demonstrated \cite{Hasko, Leti} and reliable Coulomb oscillations have been obtained up to room temperature. However, the repeatability of the process has not yet been demonstrated at the industrial scale. Still, some of these structures have been proposed as a basic element in quantum information architectures \cite{Betz}. 

In contrast, doped devices are often overlooked despite offering a reduction in the number of processing steps by avoiding the use of top- or back-gates. This lack of interest is partly due to the possibility of Anderson localisation \cite{Anderson}  as well as the randomness in dopant positions and ionisation energies that contribute to the $1/f$ noise and decrease the detection efficiency. In this type of architecture downscaling is challenging due to the high dopant density and the difficulty in realising a dopant free-tunnel barrier. However, in devices exceeding several tens of nanometers, the combination of a large number of dopants, disorder and electron interactions leads to the formation of an electron glass. The use of some of its properties, in particular, the long relaxation time $T_1$ and  charge rearrangement allow isolating electrostatically a single donor from a large number of dopants. This can be achieved by modifying the electrostatic potential of the quantum dot at room temperature and then altering charge relaxation mechanisms such as such as the neutralisation of ionised donors while the device is cooled down to the lowest temperatures. To this end, we have patterned a 60 nm diameter single-electron transistor from a highly phosphorous doped silicon-on-insulator (SOI) material using a standardised fabrication process that is described elsewhere \cite{fabrication}. The dopant concentration is about $3 \times 10^{19}$ cm$^{-3}$ giving an average donor separation of about 2 nm. The various elements of the devices are defined by etching the SOI down to the underlying silicon oxide in selected areas leaving side gates capacitively coupled to the quantum dot and tunnel barriers forming at constriction points (Fig. 1a). An additional gate connected to a double quantum dot is patterned in the same way.

A positive source-drain bias $V_{\tiny{\textup{SD}}}$ larger than the donor ionization energy ($eV_{\tiny{\textup{SD}}} \gg 45\,$meV) is first applied to the device at room temperature, allowing the creation of a steep potential profile across it without affecting its operability or its performance. The device is then slowly cooled down to the base temperature.

In such devices, the conductivity and its temperature dependence are mostly determined by the narrowest parts of the device, e.g. the constrictions and the quantum dot. By decreasing the temperature, the states located at the edge of the structures start localising electrons first, creating electrostatic tunnel barriers preferentially at the constriction locations. The effect mainly results from the presence of high-density defects on non-(100) surfaces, dielectric screening \cite{dielectric-screening1, dielectric-screening2} and the formation of $\textup{P}\textup{O}_{3-x}^-$ species within 10 nm of the edges \cite{PO3 compounds}. These states do exist in nanoscale structures even when highly doped \cite{device, IDQD}. The sidewall localisation and depletion are then responsible for the decrease in conductivity with temperature and the insulating behaviour observed in the device (Fig. 2).

Under this condition, a charge imbalance appears in the quantum dot with donors at the centre and the source-side of the island ionising and creating $P$ positive centres while a small number $N \ll P$ of electrons accumulate at the edge of the dot near the drain barrier (Fig. 1d). Raising $V_{\tiny{\textup{SD}}}$ allows increasing the electron tunnelling rate but also facilitates the higher energy confined electrons to escape to the drain contact due to the increased energy windows for conduction, thus decreasing the number of accumulated electrons. 

Consequently, most donors remain ionised during the cooling process providing that the flow of electrons through the dot and consequently the tunnelling rate through the device remains steady. This is achieved by monitoring the current $I_{\tiny{\textup{SD}}}$ and maintaining it to the room temperature value by continuously adjusting the source-drain bias $V_{\tiny{\textup{SD}}}$ during the cooldown process accordingly. Maintaining donors ionised during this process is essential as it prevents early donor neutralisation in the high-temperature range.

Once at the base temperature, the bias is slowly decreased to zero. Electrons are then allowed to recombine with ionised donors on the source side by flowing both from the source lead at a rate $\it{\Gamma}_{\tiny{\textup{S}}}$ and from the high electron density region near the drain barrier at $\it{\Gamma}_{\tiny{\textup{A}}} \ll \it{\Gamma}_{\tiny{\textup{S}}}$ (Fig. 1e). At this stage, the charge dynamic in the quantum dot becomes more complex and is driven by the difference in tunnelling rates and electron interaction. Indeed, one should notice the variation of conductivity in temperature indicates the presence of correlated hopping, e.g. Efros-Shklovskii type variable range hopping, in a large range of temperatures. The small deviation from the expected $\sim T^{-1/2}$ law \cite{ES} has already been observed in silicon MOS devices and can easily be explained by partial screening in the structure due to weak disorder, e.g. the Coulomb gap is getting partially filled \cite{Ferrus}. If charge rearrangement is done with next neighbour donors with high-level energy difference at high temperatures, at low temperatures, hopping preferentially occurs between donor pairs well separated spatially but with a small difference in energy. This behaviour demonstrates the role of both electron-electron interaction and disorder during the cooldown. Consequently, electron dynamics are governed by the one of an electron glass \cite{glass1, glass2}(Fig.2), with shallow ionised donors filled first while occupancy of deeper donors remains unsettled for much longer times. Localisation and trap density are both defined by the device structure and intrinsic randomness due to the fabrication process, including doping. Because of the spatial location of the traps in the structure, e.g. deep and shallow levels, donors at the centre of the quantum dot tend to be neutralised first and edge states at last.Traps located in the transport pathway near or at the tunnel barrier are less susceptible to electron screening due to the presence of the tunnel barrier and so, are filled at the latest. In the absence of sufficient thermal energy, traps are filled at timescales much longer than potential fluctuations' and long-term electrostatic deformation is present in the system \cite{electron glass}. This implies that already ionised donors will remain ionised for a significantly long time allowing the creation of an additional electrostatic tunnel barrier between the dot and the drain tunnel barrier with a height higher than $k_{\tiny{\textup{B}}}T$ (Fig. 1c). The width and height of the newly created barrier will depend on the competition between $\it{\Gamma}_{\tiny{\textup{S}}}$ and $\it{\Gamma}_{\tiny{\textup{A}}}$ as well as the ratio $N/P$. The later defines the average settling time that is proportional to the binomial coefficient $C_P^N \propto 1/\it{\Gamma}_{\tiny{\textup{A}}}$.

\begin{figure}
\begin{center}
\includegraphics[width=86mm, bb=0 0 575 345]{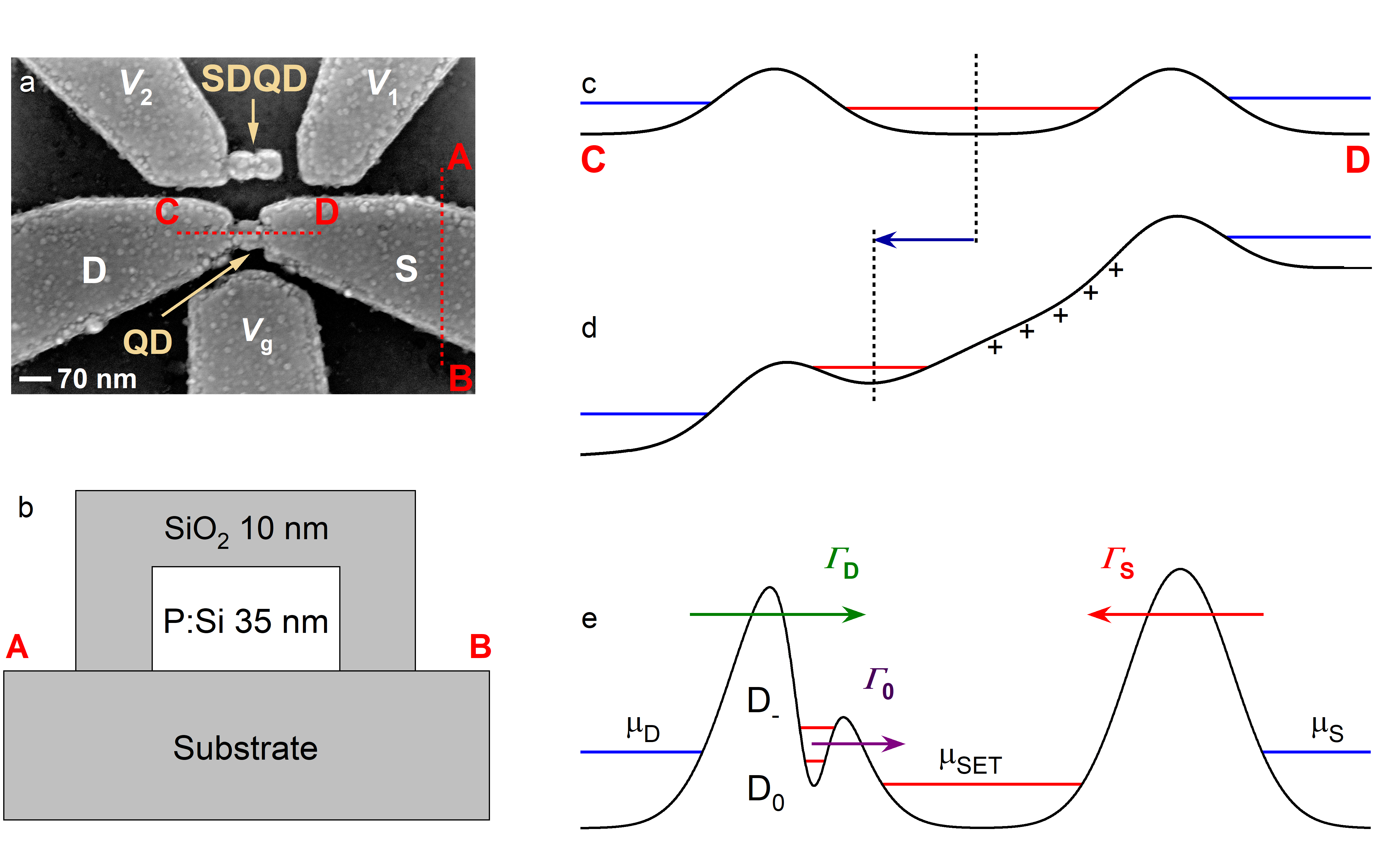}
\end{center}
\caption{\label{fig:figure1} \textbf{a} SEM image of the device used in the experiments with the quantum dot connected to source and drain leads (QD, bottom) as well as the semi-connected double quantum dot (SDQD, top) with its capacitively-coupled gate $V_{\textup{1}}$ and tunnel-coupled gate $V_{\textup{2}}$}. \textbf{b} Transverse structure of the device across A-B. \textbf{c} Energy profile along C-D at low temperatures when the cooldown is performed without biasing (all connections grounded), \textbf{d}, energy profile at intermediate temperatures for a biased cooldown and \textbf{e}, at low temperatures after the creation of the electrostatic tunnel barrier isolating the donor.
\end{figure}

\begin{figure}
\begin{center}
\includegraphics[width=86mm, bb=0 0 658 485]{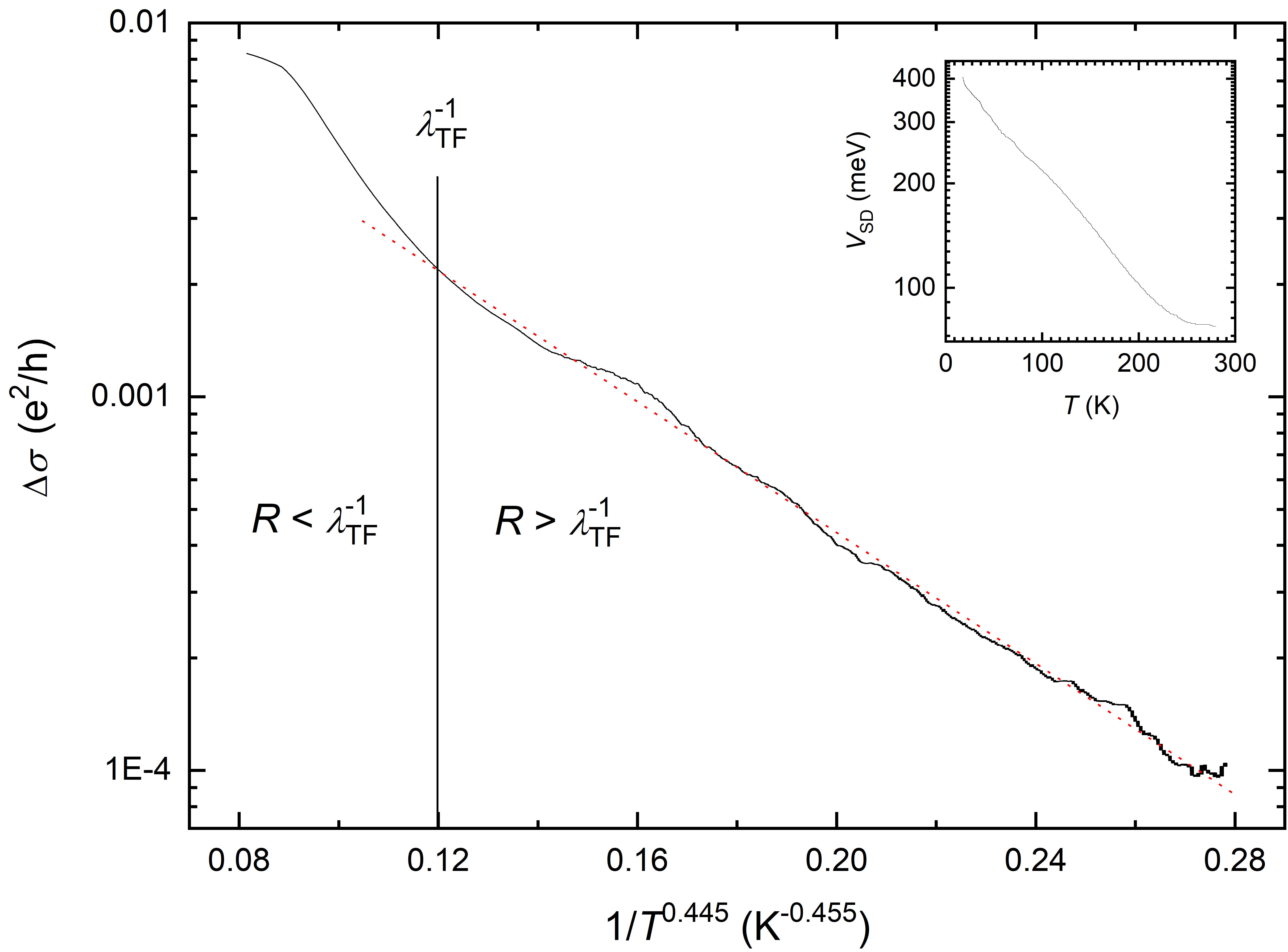}
\end{center}
\caption{\label{fig:figure2} Variation of the conductivity in temperature showing a screened correlated hopping conduction mechanism,e.g $\Delta \sigma \propto \textup{exp} (-(T_{\textup{ES}}/T)^{1/2})$, where $T_{\textup{ES}}$ is a characteristic temperature. Inset shows the variation of the applied source-drain bias during cooldown with $V_{\tiny{\textup{SD}}} \gg E_{\tiny{\textup{C}}}$, with $E_{\tiny{\textup{C}}}$ the quantum dot charging energy. The $T^{-1/2}$ law demonstrates the presence of electron-electron interaction in an insulating system compatible with $R > {\lambda_{\tiny{\textup{TF}}}^{-1}}$ with $\lambda_{\tiny{\textup{TF}}}^{-1}$ the Thomas-Fermi screening length and $R$ the hopping length \cite{ES, Ferrus}.}
\end{figure}

The resulting device structure is then similar to a donor-dot system in series \cite{Donor-dot}. Such a hybrid system has now been widely studied in particular for its ability to act as a spin readout device \cite{spin-readout} or memory in quantum information applications \cite{memory2}. However, its realization has mostly been uncontrolled with accidental trapping occurring at the tunnel barriers, unlike the proposed method.

\subsection{Device characterization}

The shape of the Coulomb diamonds following a strong bias cooldown (Figs 3a, b) is quite distinct from the one obtained on the same device but following normal cooling conditions, e. g. all connections grounded (Fig. 3c). The most noticeable difference is the appearance of large diamonds indicating the presence of impurities in the conduction path (Fig. 3a). The standard state transition from the ionised donor state $D^+$ and the singly occupied donor state $D^0$ is also clearly observable at $V_{\tiny{\textup{g}}} \sim 3.5$ V. However, there is a set of unusual characteristics. None of these features were observed in the case of the unbiased cooldown even at higher source-drain biases.

\subsubsection {Main quantum dot}

The first is the presence of low intensity, periodic and well reproducible conduction lines at the edge of the diamonds as shown on high-resolution scans (Fig. 3b). The periodicity in gate voltage $\Delta V_{\tiny{\textup{g}}}\sim $ 75 mV and the gate capacitance $C_{\tiny{\textup{g}}} = e/\Delta V_{\tiny{\textup{g}}} \sim $ 2.2 aF are similar to the ones obtained under normal conditions ($\Delta V_{\tiny{\textup{g}}}\sim $ 78 mV and $C_{\tiny{\textup{g}}}\sim $ 2.1 aF) (Fig. 3c) which points toward the main quantum dot as being the origin of these features. In the absence of an active donor in the barrier, the corresponding lever arm was found to be $\alpha_{\tiny{\textup{g}}}\sim $ 0.052 giving a charging energy of 4.1 meV. The corresponding overall dot diameter was estimated to be $D = e/(2 \pi \epsilon \epsilon_0 \alpha_{\tiny{\textup{g}}}\Delta V_{\tiny{\textup{g}}}) \sim $ 72 nm with a 58 nm of doped silicon surrounded by a 7 nm silicon oxide. Estimations were based on an effective permittivity $\epsilon = 9.9$ for the dot structure (see Suppl. II). These values were in excellent agreement with the dot diameter observed after the electron beam lithography but before oxidation of $\sim 60$ nm as well as a target protective oxide of 10 nm.

However, in the presence of a dopant in the barrier, the lever arm is found to be significantly larger ($\alpha_{\tiny{\textup{g}}}\sim $ 0.113) indicating a significant decrease in the capacitance of the drain. This can only occur if the main transport mechanism is governed by a donor in series with the quantum dot with a much weaker coupling capacitance compared to the drain lead. In that specific case, the estimation of the dot diameter from the total capacitance is no longer valid.

\begin{figure}
\begin{center}
\includegraphics[width=86mm, bb=0 0 826 713]{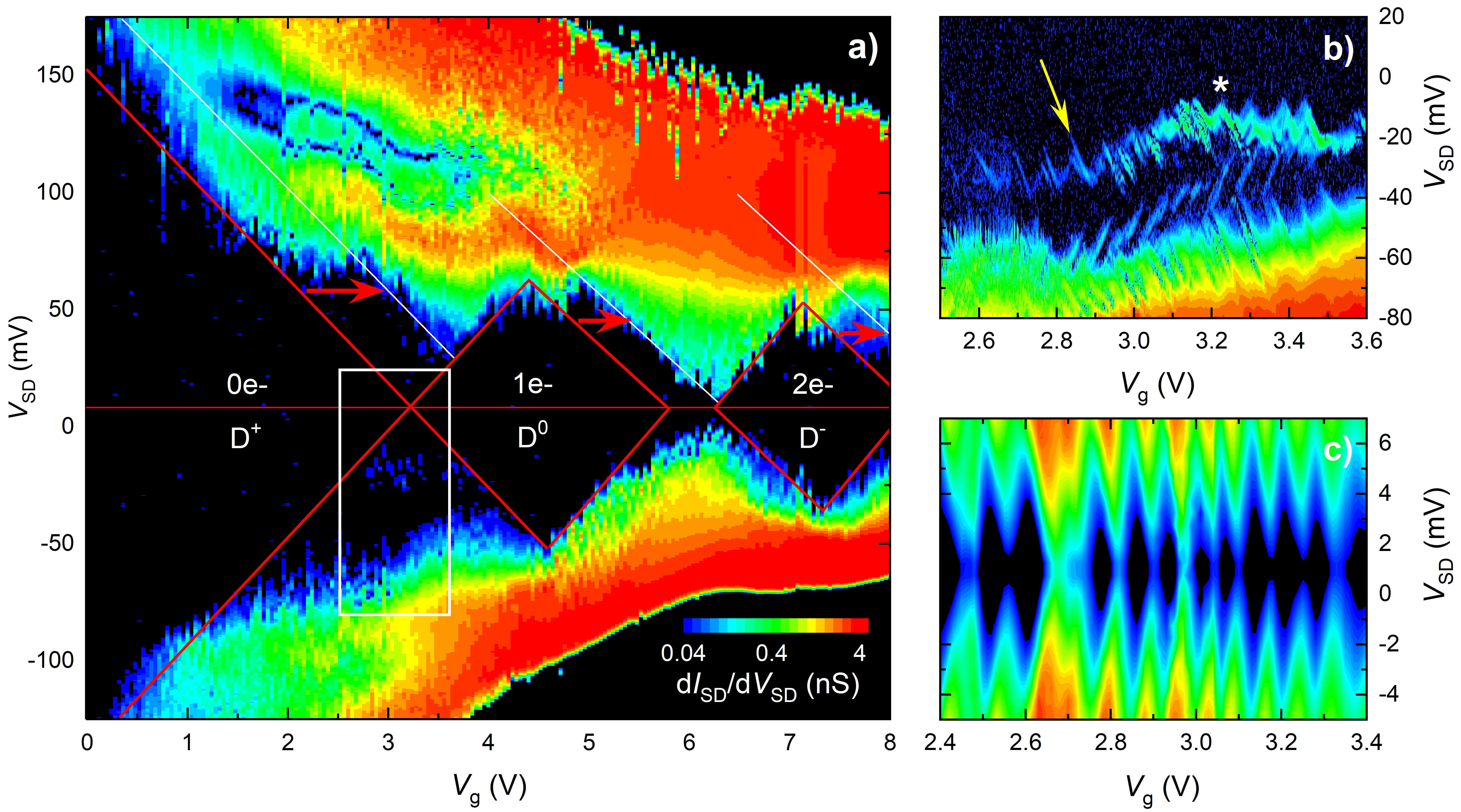}
\end{center}
\caption{\label{fig:figure3} a) Amplitude of the differential current showing coulomb diamonds at 300 mK following a cooldown under the strong bias condition b) Detailed region of figure a) around the $D^+ - D^0$ state transition (white box), showing fine features resulting from the main quantum dot, c) Coulomb diamonds at 300 mK following a cooldown under normal conditions, e.g all contacts grounded, showing the SET operating as a single quantum dot.}
\end{figure}

\subsubsection {Single donor and ionised trap}

The second effect relates to the appearance of shifts in the large Coulomb diamonds both for the D$^+$, $D^0$ and doubly occupied D$^-$ states (marked by arrows in figure 3a). Understanding the origin of these shifts is important to precisely determine the donor energies as well as to adequately perform the charge detection carried out in the next section. Such shifts have already been observed in FinFets where two donors have been diffusing from the contacts into the channel region \cite{Shifts}. In this case, the transport was found to be mainly dominated by a single donor (D1) whereas an ionised trap (D2) had its occupancy modified under certain voltage conditions. A similar situation happens here. 

Charging energies $E_{0}$ and $E_{-}$ for both the D$^0$ and D$^-$ states of D1 can be estimated when the trap D2 is ionised by either measuring the size of the diamonds along the $V_{\tiny{\textup{SD}}}$ axis or the $V_{\tiny{\textup{g}}}$ axis by converting voltages into energies using the measured lever arm $\alpha_{\tiny{\textup{gD}}} \sim$ 0.022. The corresponding D1 level energies are found to be E$_0 \sim 67$ meV and E$_- \sim 57$ meV, consistently by both methods. These values are far greater than the one expected for isolated donors in bulk silicon \cite{donor-charging} but have already been experimentally observed in silicon quantum dots \cite{Donor-dot}. However, these differences can easily be explained by the device structure and doping. In nanostructures, the presence of interfaces reduces the extension of the wavefunction and in the case of the device studied, the bound state wavefunction is elongated in the current direction, e.g., along the source-drain axis, while being reduced in the transverse direction \cite{extension}. This increases significantly the separation between the ground and excited levels, and so between the D$^0$ and D$^-$ states. This is particularly true as the isolated dopants are located at the constriction, e.g., the narrower part of the device. Also, energy levels are sensitive to the electrostatic environment, in particular, the electric field at boundaries which leads the effective surrounding permittivity to be renormalised and the localisation to be enhanced. We did estimate the effective permittivity to be $\epsilon \sim$ 9.9 in the previous sections \cite{dielectric-screening1, dielectric-screening2, permittivity}. This leads naturally the charging energy of the D$^0$ states to be raised to 63 meV without any other adjustment.

When D2 is occupied by an electron, the local electric field at D1 is modified and its energy levels shift by the screened Coulomb interaction between the tunnelling electron and the electron localised on D2. This allows estimating an average distance between the two donors in the tunnel barrier of $d \sim $ 1.4\,nm, a value close to the effective Bohr radius in silicon (see. Suppl. III). Such a cluster state has already been identified in highly-doped devices by Kelvin probe force microscopy (KPFM) \cite{cluster}. The shifts observed on the D$^0$ and D$^-$ states are a direct consequence of this interaction and are related to the shift of the D$^+$ state.
These features are well described quantitatively and qualitatively by simulations including a donor and an ionised trap near the drain barrier and in series with the main quantum dot (see. Suppl. III, IV and V)

\subsubsection {Additional features}

\begin{figure}
\begin{center}
\includegraphics[width=86mm, bb=0 0 768 509]{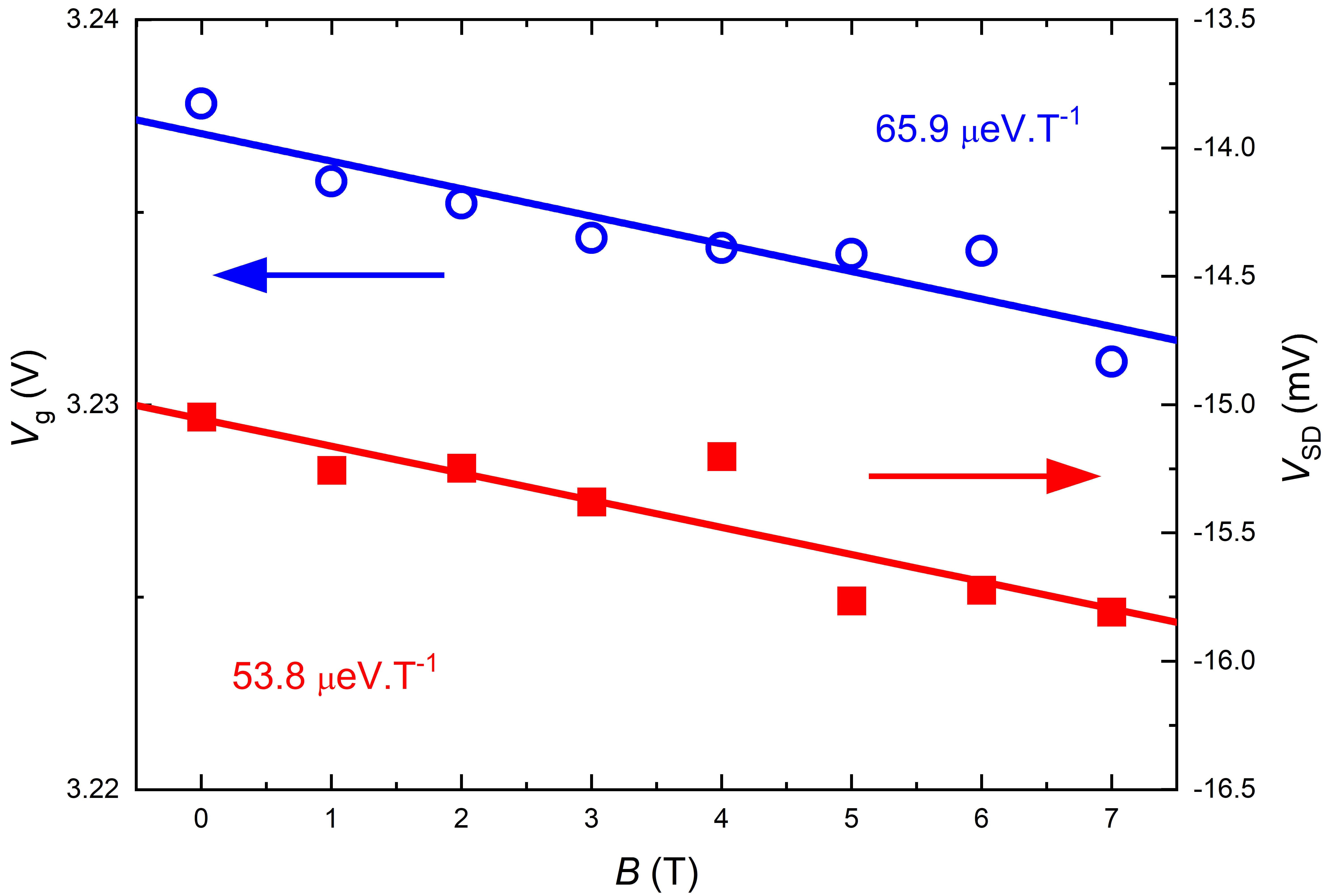}
\end{center}
\caption{\label{fig:figure4} Displacement in $V_{\tiny{\textup{SD}}}$ and $V_{\tiny{\textup{g}}}$ of a typical feature (white star) at the $D_+$-$D_0$ transition as a function of the magnetic field applied perpendicularly to the device.}
\end{figure}

Additional features are present in the data. In particular, at large $\mid V_{\tiny{\textup{SD}}} \mid$ and 1\,V $< V_{\tiny{\textup{g}}} <$4\, V, one can observe the presence of negative differential conductivities (NDC) with isoenergy lines following the edges of Coulomb diamonds (Fig. 3 and Suppl. VI). Such behaviour has already been observed and previously discussed in detail in similar doped silicon devices \cite{NDC1,NDC2}. In the latter case, the NDC resulted from the modification of the shape of the tunnel barrier and so, of the tunnelling rates due to the presence of ionised donors near or at the barriers. This shows the NDC is an intrinsic property of doped devices rather than an effect of the cooldown process. However, the distribution of dopants near the barrier, and so, the shape of the tunnel barrier favour the formation of isolated but transport active donors.

We also observed the absence of conductivity at negative source-drain bias and 2\,V $ < V_{\tiny{\textup{g}}} <$ 4\,V, in the region where D2 is singly occupied and a second electron is tunnelling through the D$^0$ of D1. Such an effect is commonly associated with spin-related phenomena \cite{spin-blockade}. Although full study of spin states and spin interaction in that device has not been carried out, this observation underlines the role of spin in that region of gate and source-drain biases. Only a small region around the transition between the D$^+$ and D$^0$ state shows active transport (2.6\,V $ < V_{\tiny{\textup{g}}} <$ 3.6\,V) (Fig. 3b). These features can all be displaced linearly by the application of a magnetic field perpendicular to the device both along the $V_{\tiny{\textup{SD}}}$ and $V_{\tiny{\textup{g}}}$ axis (Fig. 4) suggesting the Zeeman effect is active in this region. Results show an average energy level displacement by $\sim 60\,\mu$eV T$^{-1}$, consistent with the expected value of $58\,\mu$eV T$^{-1}$ for a 1s state electron with an effective Land$\acute{\textup{e}}$ g-factor of 2 \cite{Zeeman}.

Consequently, these features around the D$^+$ and D$^0$ transition well correspond to the state where an electron is tunnelling via the available lowest level of the D$^0$ state of the donor D1 whereas the trap D2 remains ionised in the D$^+$ state. We will use the feature at $V_{\tiny{\textup{g}}} \sim 2.8$ V (yellow arrow in Fig.3b) for detection purposes in the next sections owing to its low noise property.

\subsection{Numerical simulations}

To assert the correctness of the proposed model, e.g. two donors in parallel with each other but in series with the quantum dot, we performed a set of numerical simulations. For this, a constant interaction model was used to solve a master equation and find both the equilibrium states of current and the charge occupancy at given external voltages (Suppl. V, \cite{CoulombTheory}).

In the model, both donors are capacitively coupled to the drain reservoir and the quantum but only one is active in transport, e.g. has a finite tunnel resistance. This is what we observe experimentally. Obviously, the arrangement of the various gates around the device implies complex capacitive couplings to each donor and the quantum dot, rendering calculations extremely time-consuming (Fig. 5a). However, it is possible simulating each individual feature observed experimentally (Sec. Device characterisation) while simplifying the problem significantly.

We first start by simulating the device for the unbiased cooling (Figs. 5a and 6c). In this case, no donors are active in transport and only a single quantum dot is present. This allows calibrating some of the capacitances, in particular, $C_{\tiny{\textup{g}}}$ the coupling capacitance between the gate lead $V_{\tiny{\textup{g}}}$ and the quantum dot, and $C_{\Sigma}$ the total dot capacitance. Best values are found to be $C_{\tiny{\textup{g}}} \sim$ 2.1 aF and $C_{\Sigma} \sim$ 34 aF, which are in the agreement with experimentally observed values.

Noticing that the periodicity of the oscillations at the edge of the diamonds is not significantly affected by the values of the gate and source-drain biases, we reduced the problem to a small quantum dot in the barrier in series with a larger one (Fig. 5b). Results shown in Fig. 6b well reproduce experimental observations for a small dot capacitance of $C_{\Sigma 0} = 2.8$ aF, and are consequently compatible with the presence of a donor.

Concerning the observed shift in gate voltage at the $D^+$, $D^0$ and $D^-$ states, we restrict this time the model to a donor and a trap, i.e. D1 and D2, both in the drain tunnel barrier (Fig. 5c). Each donor is modelled by discrete energy levels on top of their capacitive potential energy. These discrete energy levels reflect the $D^0$ and $D^-$ donor charge states. When a repulsive screened Coulomb interaction between the charges held by the donors is added, the simulations reproduce well the observed features in position and amplitude (Fig. 6a). 
 
\begin{figure}
\begin{center}
\includegraphics[width=86mm, bb=0 0 450 250]{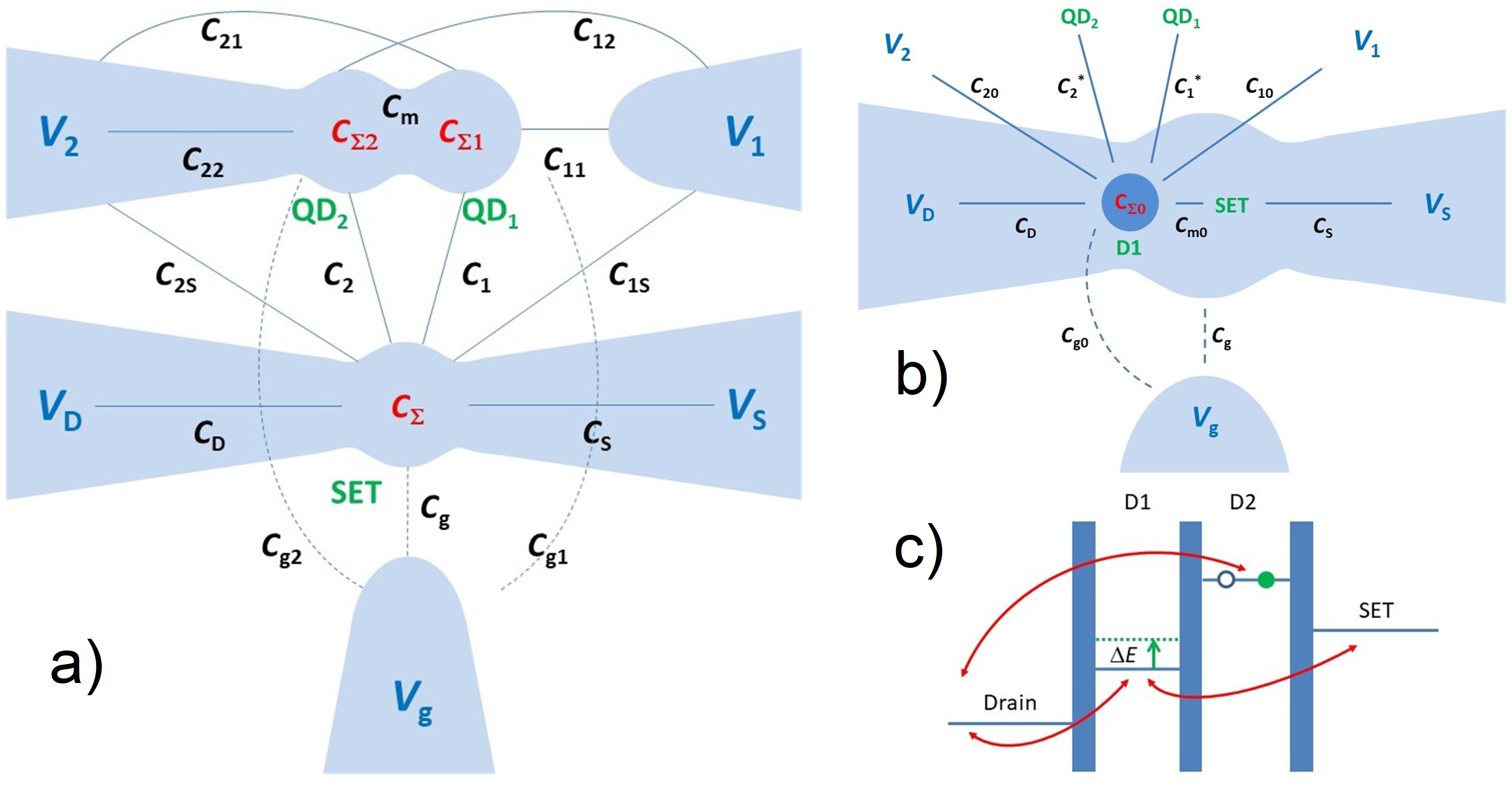}
\end{center}
\caption{\textbf{a} Capacitance model for a semi-connected double quantum dot and its charge detector represented here in its simplified version by a single electron transistor, \textbf{b} Closer view of \textbf{a} (white box) of the capacitance associated with the active donor D1, \textbf{c} Schematics of the energy levels and tunnelling processes involving the two donors. Tunnelling from and to the drain contact and the SET are allowed via D1. However, due to the energy difference, such a process is not allowed in D2 which can only be charged and discharged via the drain.}
\end{figure}

\begin{figure}
\begin{center}
\includegraphics[width=86mm, bb=0 0 1042 563]{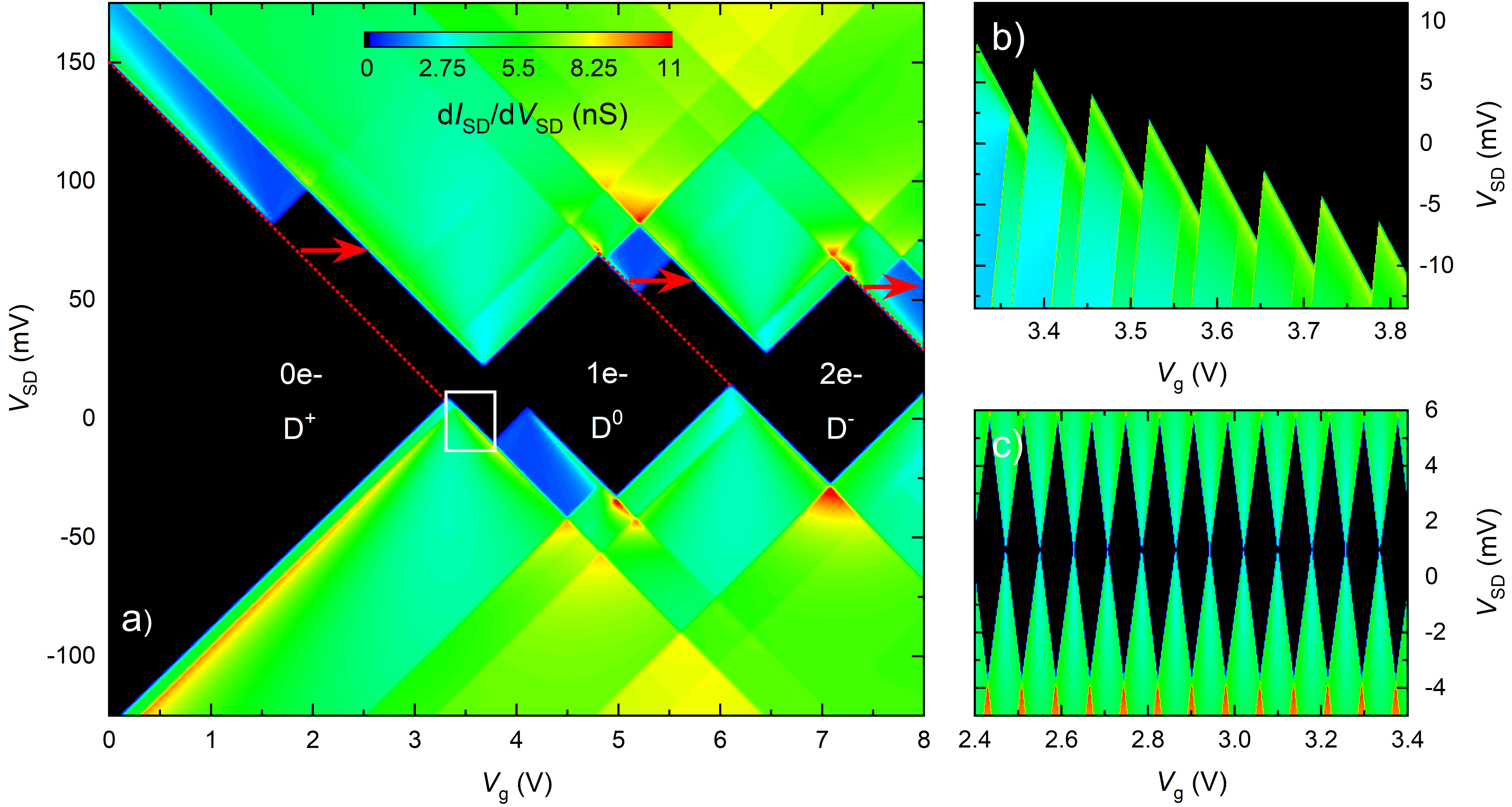}
\end{center}
\caption{\textbf{a} Simulated Coulomb diamonds showing, the $D^+$, $D^0$ and $D^-$ with shifts induced by the presence of the second impurity in the tunnel barrier. \textbf{b} Details at the edge of the diamond showing the effect on the quantum dot when in series with the donor (white box area in \textbf{a}). \textbf{c} Simulation of the Coulomb diamonds in the unbiased cooldown case, e.g. in the absence of donors in the barrier.}
\end{figure}

\subsection{Application: using the D$^{0}$ state as a charge detector}

The previous results allow the possibility of using the donor as a detector rather than the main quantum dot as this would normally be the case in most experiments. Such a scheme would present a significant advantage in terms of detection efficiency. In a hydrogen-like donor, only two main states are allowed: the D$^0$ and D$^-$ states. If the detection point is chosen to be at the D$^+$-D$^0$ transition then tunnelling events are restricted to the D$^0$ state only as, at low temperatures, the D$^-$ is energetically not accessible without significantly modifying $V_{\tiny{\textup{g}}}$ and $V_{\tiny{\textup{SD}}}$. Consequently the current will only be able to flow selectively when electrons tunnel into the ionised donor at the energy level corresponding to the $D_0$ state.

The difference in measured currents between the blocked and allowed tunnel events then makes the detection of a capacitively coupled structure easily observable. On the contrary, in a quantum dot, multi-tunnelling processes are allowed, including inelastic ones like cotunnelling. Consequently, the detection efficiency is reduced to a fraction of the one of a single donor. Additionally, the donor levels are known to be very sensitive to the electric field. Any modification of the electrostatic potential would then shift the levels of the donor and the detection point significantly, improving detection (Suppl. VII, \cite{Fernando, Schaal}.

In the present device, a semi-connected double quantum dot (SDQD) is capacitively coupled to the SET and the isolated donor (Fig. 1a). In the SDQD one dot (QD2) is connected via a tunnel barrier to a side gate ($V_2$) that acts both as an electron reservoir and a gate, whereas the second (QD1) is capacitively coupled to QD2 and the gate $V_1$. To properly use the donor as a detector, a double gate compensation has to be performed. Such a technique, mostly known as 'virtual gating', consists in modifying simultaneously a pair of gate voltages to eliminate the influence of the main dot (SET) on the detection. The charge diagram of the double dot is then obtained by sweeping the gate $V_2$ and stepping the gate $V_1$ while keeping the detection point at the D$^0$ level stable. This implies modifying the SET gate voltage $V_{\tiny{\textup{g}}} = V_{\tiny{\textup{g0}}} +\alpha V_1+ \beta V_2$ where $\alpha$ and $\beta$ are the compensating coefficients and $V_{\tiny{\textup{g0}}}$ is the initial value of the gate voltage. As the detection point is at the edge of the D$^+$-D$^0$ transition, the source-drain bias $V_{\tiny{\textup{SD}}}$ has also to be modified in the same way by following the compensation line in the Coulomb peak (direction of the yellow arrow in Fig.3b). Optimal detection is obtained with $\alpha \sim -0.284 $ and $\beta \sim -0.373$.

The resulting stability diagram between $V_1$ and $V_2$ shows clear characteristics of a weakly coupled double quantum dot structure, in particular, the double periodicity ($\Delta V_1$ along $V_1$ and $\Delta V_2$ along $V_1$) and transition lines nearly parallel to the gate axis Fig. 7a). By averaging across the measured plot in Fig. 7, one obtains $\Delta V_1 \sim $ 219 mV and $\Delta V_2 \sim $ 599 mV. Periodic current peaks along $V_1$ are consistent with electrons tunnelling to the dot closer to the $V_1$ gate. The absence of sharp edges in the transitions is explained by the small difference in capacitances between the individual dots of the double quantum dot and the donor, i.e. $C_{\tiny{\textup{1}}} \sim C_{\tiny{\textup{2}}}$ (Fig. 7b and Suppl. IV). On the contrary, $V_2$ both acts as a gate and an electron reservoir to the double dot structure. This gives the possibility for the electrons to tunnel out of the double dot, consequently causing a larger change in the conductivity at the donor site and so, abrupt jumps in $V_2$ (Fig. 7d). This allows estimating the gate capacitances for $V_1$ and $V_2$, e.g, $C_{11} \sim e/\Delta V_1 \sim 0.73$ aF and $C_{22} \sim e/\Delta V_2 \sim 0.27$ aF respectively.

We also notice a shift $\delta V_1$ in the pattern along $V_1$ when $V_2$ is increased, e.g. when QD2 acquires an additional electron (red arrow in Fig.7 a). In this case, adding an electron to DQ1 from the reservoir now requires overcoming the interaction energy between the two dots which leads to a shift in $V_1 \sim C_{\tiny{\textup{m}}}/C_{\Sigma1}C_{\Sigma2}$. 

Bistable regions are present at around $V_2 = $ 1.25 V, 2 V and 2.75 V (Fig. 7a). One has to remind that if the value of the conductivity depends on the tunnelling events at the detector, the location of the electronic transitions in the compensated diagram result from electron tunnelling in the double quantum dot structure. When $V_2$ increases three tunnelling events are possible, i) an electron can enter QD2 directly, (ii) an electron can enter QD1 by overcoming the interdot coupling energy, (iii) an electron enter QD2 while another electron tunnel between QD2 and QD1. These processes being random and the measurement being time-averaged ($t_{\tiny{\textup{measure}}} \gg t_{\tiny{\textup{tunnel}}}$), this phenomenon is reflected through the observed bistability Fig. 7a and c). By the direct measurements of the periodicities in $V_1$ and $V_2$, the shift $\delta V_1$ in $V_1$ and the width of the bistable region $\delta V_2$ in $V_2$ (Fig. 7a), it is possible to extract capacitance parameters, in particular the coupling capacitances between the donor and the gate $V_2$, e.g. $C_{2}^{*} \sim 0.12$ aF, and between the donor and the gate $V_1$, e.g. $C_{1}^{*} \sim 0.27$ aF, as well as the coupling capacitance between QD1 and QD2 $C_{\tiny{\textup{m}}}^{'} \sim 0.14$ aF (Suppl. IV). Such a low value for $C_{\tiny{\textup{m}}}^{'}$ well explains the approximate square shape of the charge stability diagram in Figure 7.

\begin{figure}
\begin{center}
\includegraphics[width=86mm, bb=0 0 803 498]{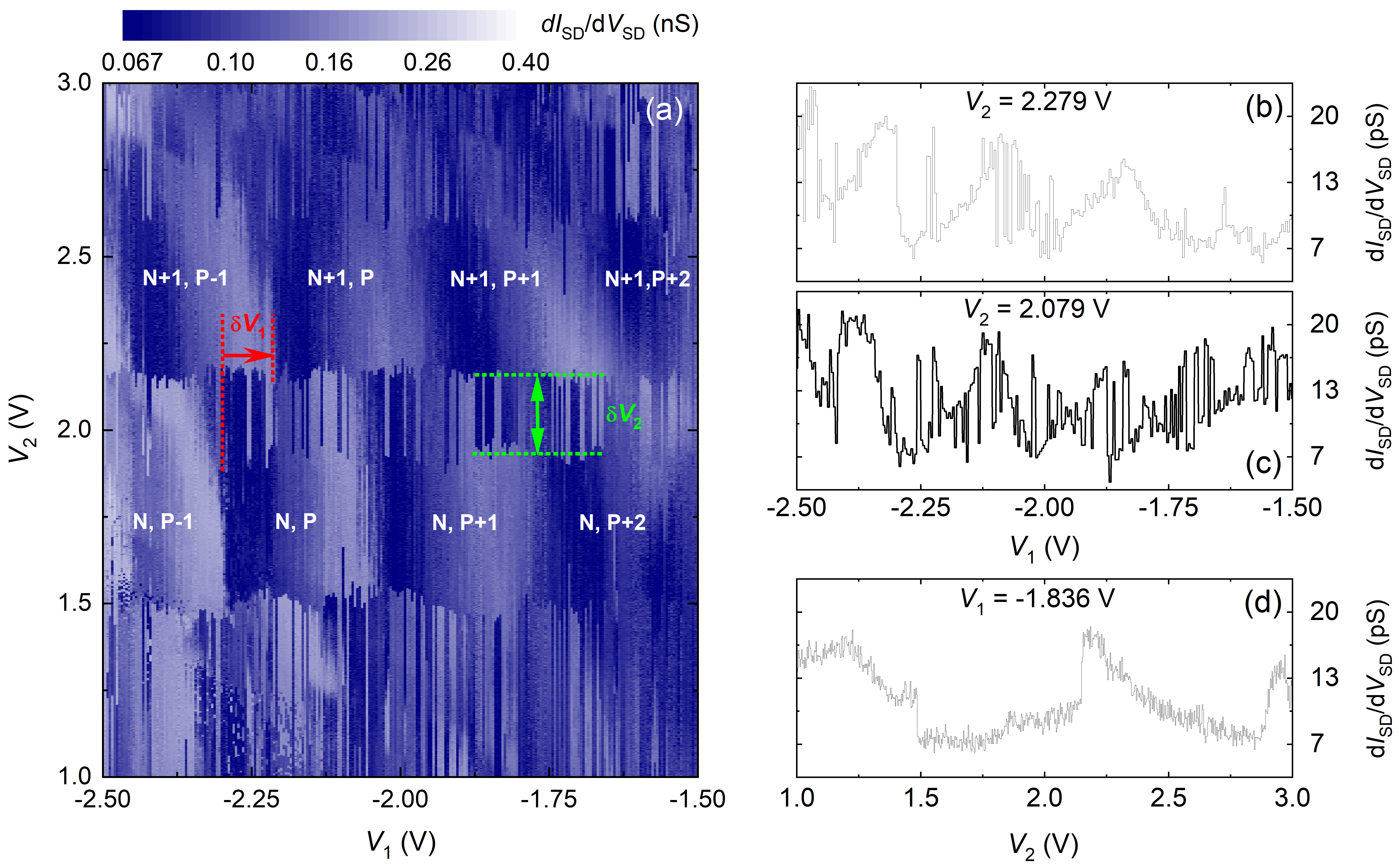}
\end{center}
\caption{\label{fig:figure7} \textbf{a} Compensated charge stability diagram for the SDQD at 300 mK, \textbf{b} Current profile along $V_1$ showing individual electron tunnelling event inside the double dot, \textbf{c} Current profile at the bi-stable region involving tunnelling events from individual dots to the gate $V_2$, \textbf{d} Current profile along $V_2$ indicating electrons tunnelling out of the double dot into the gate reservoir.}
\end{figure}

\section{Conclusions}

We have shown that a single donor could be electrostatically isolated from a large ensemble by making use of the intrinsic glassy behaviour of a doped semiconductor device and reshaping the quantum dot potential. The creation of an extra potential barrier allows the realisation of a donor-quantum dot hybrid system. By adjusting the gate and source-drain voltages, the electrostatic influence of the quantum dot can be cancelled out and one can use the D$^{0}$ state of the donor to turn the device into a charge detector formed from a single atom and map the charge states of a nearby but capacitively coupled double quantum dot. This method is largely applicable to all doped semiconductor quantum dots and provides a reliable, low-cost and fast way of realising single-donor structures that could be utilised in semiconductor quantum technology applications such as charge detectors. In principle, such detectors could be used as part of the measurement stage of a quantum processor.

\section{Acknowledgement}

This work was supported by Project for Developing Innovation Systems of the Ministry of Education, Culture, Sports, Science and Technology (MEXT), Japan and by Grants-in-Aid for Scientific Research from MEXT under Grant No. 22246040. T. Ferrus is grateful to Prof. Sir M. Pepper from the University College London for discussion and feedback on the manuscript, as well as to Prof. Chris Ford from the University of Cambridge, for software and driver development and Dr Aleksey Andreev for initial simulations. Finally, the authors would like to thank Dr. Alessandro Rossi from the University of Strathclyde for useful discussions and preliminary assessment of the device.


\begin{references}
\bibitem{device} Doping concentration is about 3 $\times 10^{19}$ cm$^{-3}$ which provides metallic behaviour in the large areas of the device, e.g. source and drain leads. However, the breaking of symmetry at the structure edges, near the Si-SiO$_2$ interface does not provide sufficient electron screening and localisation occurs.
\end{references}
\end{document}